\documentclass[runningheads]{llncs}
\usepackage{booktabs}
\usepackage{rotating}
\usepackage{multirow} 
 \usepackage{array} 
 \usepackage[export]{adjustbox}
 \usepackage{float}
\usepackage[T1]{fontenc}
\usepackage{threeparttable}
\usepackage{subcaption}
\usepackage{graphicx}
\usepackage{graphicx}
\begin{document}
\title{Digital Weight Management Interventions: A review of commercial solutions and survey analysis of user needs}
\author{Suncica Hadzidedic\inst{1}\orcidID{0000-0001-9026-8737} \and
Jingyun Wang\inst{1}\orcidID{0000-0001-9325-1789} \and
Victor Elijah Adeyemo\inst{1,2}\orcidID{0000-0002-8398-3609} \and
George Sanders\inst{2}\orcidID{0000-0001-6458-7312} \and
Grant Westermann\inst{2}
}
\authorrunning{F. Author et al.}
\institute{Durham University, Durham DH, UK \\
\email {\{victor.e.adeyemo, suncica.hadzidedic, jingyun.wang\}}@durham.ac.uk\\ \and
MoreLife (UK) Ltd, Leeds, LS2 3AA, UK\\
\email{\{george.sanders,grant.westermann\}@more-life.co.uk}
}
\maketitle              % typeset the header of the contribution
\begin{abstract}
Obesity is a global health challenge. According to the World Health Organization (WHO), between 1990 and 2022, adult obesity more than doubled. Weight management interventions (WMIs) support individuals in achieving and maintaining a healthy weight through dietary guidance, physical activity promotion and behavioural counselling. However, traditional WMIs often have limited accessibility. Digital WMIs or DWMIs are delivered via websites or smartphone applications and provide scalable and cost-effective alternatives. However, user needs for digital services and their prevalence in the existing commercial solutions remain underexplored. Hence, our study systematically identified 26 commercial DWMIs to identify their features, services, and data collection practices. Additionally, we performed a user needs analysis by recruiting 207 individuals involved in a real-life WMI. Our findings indicated that DWMIs integrated self-monitoring, goal setting, and behaviour change strategies, yet lack social support, virtual reality applications and adaptive personalisation. WMI clients prefer smartphone Apps and fitness trackers for tracking weight management progress and have varying levels of comfort in using digital resources. The presented results serve as recommendations for future directions in the design and implementation of services for DWMIs. 

\end{abstract}
\section{Introduction}
Obesity and weight problems have been increasing globally for decades, and according to the World Health Organisation 2.5 billion adults are overweight \cite{WHOObesity}. Between 1990 and 2022, the prevalence of obesity among children and adolescents quadrupled, while it more than doubled among adults, rising from 7\% to 16\% \cite{WHOObesityRate}.  Obesity is a chronic non-communicable disease, which increases the risk of comorbidities, e.g., type 2 diabetes and heart disease \cite{Apolzan,Moravcová} and is associated with mental health issues \cite{Avila}, e.g., depression. Consequently, effective weight management is vital.

Weight management interventions (WMIs) refer to programmes that assist individuals manage weight, build healthy lifestyle habits and maintain weight loss \cite{NHSObesity}. Traditional WMIs are delivered in-person \cite{NHSObesity} via healthcare providers, structured weight-loss clinics, or community-based programmes. However, this often requires significant resources \cite{Fan}, which can limit accessibility.
Given these challenges, digital weight management interventions (DWMIs), which are delivered through websites or smartphone applications, emerged as scalable, accessible and cost-effective solutions \cite{Cheah}. The popularity and adoption of DWMIs was driven by the increase in smartphone users, estimated to be 7.41 billion globally in 2024 \cite{mobileUsers}. Prior studies reviewed some of the available DWMIs, e.g., Noom \cite{Behr}, Vitadio \cite{Moravcová}, and Weight Watchers \cite{Apolzan,Pagoto}. However, a comprehensive understanding of services and features delivered and accessible via DWMIs remains underexplored. Moreover, while user-centred design \cite{Sugandh} can foster usability and effectiveness of technology, user needs for DWMI services have not been adequately assessed in related work.

Our study therefore provides a comprehensive analysis of existing commercial DWMIs and preferences of actual users, with the aim to identify directions for future design and implementation of digital interventions in the weight management domain. We contributed to this aim by performing the tasks below:  
(i) a systematic review of commercial DWMIs to identify the prevailing features or services, and type of user data collected; and
(ii) a survey of user needs and its analysis, with individuals involved in a real-life WMI, who identified preferred digital approaches based on their experience in WMIs.
Both tasks were performed independently, then we synthesised the obtained results.

\section{Literature review}
DWMIs have gained significant traction \cite{Cheah}, leveraging technology to enhance user engagement, behavioural modification and self-monitoring. Studies have evaluated (and reported features that affect) the effectiveness of DWMIs. In \cite{Moravcová}, the Vitadio app was used for a 6-month structured and personalised digital lifestyle modification and self-management programme. They reported a reduction in participants' body weight and fat. Similarly, \cite{Silberman} found that a DWMI offering one-on-one health coaching and engagement features (e.g., messages and tracking) over 12 months led to significant weight loss. Moreover, \cite{Pagoto} showed that active participation in virtual workshops and private community engagement improved weight loss and programme satisfaction in a DWMI with integrated WeightWatcher. This aligns with findings from \cite{Kupila}, who examined the Health Weight Coaching (HWC) programme and showed that frequent weight reporting and higher initial BMI were significant factors for programme adherence and weight loss. Other studies, e.g. on NoHow toolkit \cite{Marques}, suggested that DWMIs with psychological strategies enhanced long-term weight management, and that self- and emotion-regulation significantly contributed to weight loss maintenance.

The integration of digital therapeutics with personalised health data is one emerging approach. The Digbi Health programme \cite{Sinha} incorporated genetic and gut microbiome profiling for personalised dietary recommendations. Similarly, \cite{Bermingham} implemented an 18-week app-based personalised dietary programme. Moreover, personalised avatars in DWMI  reportedly contributed to better weight outcomes \cite{Horne}. Recently, AI-driven digital interventions have also gained prominence. For example, \cite{Graham} used conversational AI and identified AI interactions as one of the key factors towards weight maintenance. 

Other studies evaluated the usefulness and user experience of DWMIs. E.g., \cite{Ghelani} reported qualitative factors influencing the use of DWMIs, such as information provision, behaviour change strategy and social engagement. Furthermore, \cite{Naabi} investigated designing sustainable DWMIs and proposed including features such as diet management, exercise, activity planning and goal setting. Overall, DWMIs' effectiveness has been explored in prior studies. However, there is a lack of studies that systematically reviewed DWMIs to identify the common and unique services and features they provide, and research on target user needs for DWMI services and features.

\section{Methods}

\subsection{Commercial DWMIs: Review methods}
We used three online platforms to identify DWMIs: Apple App Store, Google Play Store, and Reddit. The Apple App Store and Google Play Store are official repositories for iOS and Android applications, respectively. Reddit, with approximately 57 million daily users as of December 2022 \cite{RedditDailyUsers}, hosts user discussions on various topics, including DWMIs. 

We applied the following steps to identify, select, download and review commercial DWMIs (Results in Section \ref{dwmi}): 
\begin{enumerate}
    \item In Reddit we applied the search string “‘weight OR obesity AND management’ apps”, which provided users posts from related subreddits:  r/diet, r/ProductivityApps, r/weightlosstechniques, r/selfimprovement, and r/loseit. We identified 26 trending DWMIs across the subreddits by reading through top users' comments.
    \item Apple App Store and Google Play Store were used to manually extract statistical data about the identified DWMIs into a Microsoft Excel file. Finally, 17 of 26 DWMIs satisfied these inclusion criteria: (a) The DWMI must be categorised as a “Health and Fitness” application in both Apple App and Google Play Store, as it is more likely to offer relevant services for weight management, (b) English must be one of the supported languages to cater to global audience, and (c) The DWMI's minimum average rating must be 4.0/5.0 on either Apple App or Google Play Store as an indicator of popularity and relevance to the target users. 
    \item The 17 selected DWMIs were downloaded from Google and Apple app stores and installed on one smartphone. Each DWMI was tested for 2 sessions within a month, to identify and review the core system features, content, and services provided.  Only free functionalities were accessed and used. 
\end{enumerate}

\subsection{Methods for user needs analysis}
In addition to the review of existing commercial DWMIs, we surveyed real-life clients, of a UK company that delivers weight management interventions, about their perceptions and needs for DWMI. The user study was reviewed and approved by the University's Ethics Committee. Convenience sampling was applied and we recruited 207 participants, who completed their weight management intervention or attended at least 6 of 12 weekly meetings. Participants' consent was obtained and they were informed of the right to withdraw. No incentives were provided.

We performed the study during May and June 2024 and used an online survey to collect data. Participants were asked 20 questions: 13 multiple choice, three on a 5-point Likert scale, and four open-ended, on topics including motivation, interests, preferred weight management tracking, preferred weight management support or resources, challenges and stressors, and demographics. The participants' data were anonymised and stored securely. 
Only two of the survey's three 5-point Likert scale questions (i.e., “How comfortable do you feel using each option for tracking your weight management progress?” and “What support or resources were/are helpful to you during the weight management programme?”) are relevant to this study and thus included. We applied descriptive statistics (1 = very uncomfortable/helpful, 2 = Uncomfortable/helpful, 3 =Neutral, 4 = Comfortable/helpful and 5= Very Comfortable/helpful) and t-tests to investigate significant differences between groups.

\section{Results} 

\subsection{Reviewed commercial DWMIs}
\label{dwmi}
The information of 17 DWMIs are presented in Table \ref{table1}, containing their availability as a smartphone application or web platform, popularity in app stores and user ratings. MyFitnessPal, WeightWatchers, and Lose It! were the top-ranked DWMIs in terms of downloads and ratings, while DietSensor is the lowest in downloads.

The DWMIs collected explicit user input data at account creation using 65 questions which were manually grouped into six categories, namely: (i) Health and medical conditions, (ii) Diet, Activity and Fitness, (iii) Person profile and lifestyle, (iv) Tracking and Consistency, (v) Users' goal setting, motivation, previous experience, barriers, and preference, and (vi) Experience, Journey and Mindset. Following the account creation process, each DWMI presented the user with a specific or holistic goal or outcome(s). The outcomes identified across the 17 DWMIs are: nutrition, exercise, fasting, weight logging and behaviour change plans. An example of a specific (nutritional) outcome is a personalised plan based on the user's profile and estimated metabolic age used to calculate daily goals that were adapted to the user's schedule offered by the Keto diet app. 
An example of a holistic (i.e. behaviour change, nutrition and fasting) outcome is a personalised daily calorie budget, meal planning and target, recipe recommendation, habit pattern analysis, and a custom dashboard theme offered by Lose it! App.

Our review indicated that eight features were implemented across the commercial DWMIs, including various logging facilities (Table \ref{table1}, numbers 9-20), reminder or notification system, achievement and reward system, food or barcode scanner, coaching and practitioner support, community support and sharing, synchronisation with external tracking apps and synchronisation with external sensing app (Table \ref{table1}, numbers 21-27 respectively). Noom, Calorie Counter, and Fastic: Fasting \& Food Tracker were the DWMIs that integrated at least seven of the eight core system features, while the Carb Manager - Keto Diet Tracker DWMI featured only the logging facilities. Logging capabilities as a core system feature differed across DWMIs. While Lose It!, Yazio, MyFitnessPal, and Carb Manager provided comprehensive tracking of weight, diet, exercise, and metabolic health, others focused on specific aspects, such as fasting or calorie intake. 

The reviewed DWMIs included six types of resources provided as services to users. 15/17 DWMIs provided recipe or food database as a service, six DWMIs offered video or audio for activities, five DWMIs provided courses, four DWMIs presented healthy lifestyle manuals, four DWMIs offered success story sharing, and only three DWMIs had podcasts (Table \ref{table1}). Personalised services, i.e., content and resource recommendations, were sporadically seen. DWMIs offered six types of recommendations (Table \ref{table1}, numbers 28-33). The recommendation level (i.e., Low, Medium and High) for each DWMI (Table \ref{table1}, number 34) was manually synthesised based on the number of distinct recommendations offered. Three DWMIs (i.e., DietSensor, Fastic, and Lifesum) provided four or more types of recommendations, therefore graded as high recommendation level. Meanwhile, other DWMIS offered at most three recommendations and were graded as low or medium recommendation levels, respectively. 

Overall, the reviewed DMWIs collected user input data via 65 questions manually categorised into six groups. Following user account creation, five outcomes made available to users, that offered either specific or holistic plans. Eight core system features were identified across the DWMIs (with 12 distinct logging facilities as one of the features), meanwhile, six types of resources and six types of recommendations were provided as services.

\begin{sidewaystable}
\rotatebox{180}{
\begin{threeparttable}
\caption{Commercial DWMIs: Statistics from app stores, common features, services, recommendations}
\label{table1}
\centering
\begin{tabular}{@{}p{2.7cm}cccccccccccccccccccccccccccccccccccccccc@{}}
\toprule
\multirow{3}{*}{DWMI} & \multicolumn{2}{c}{Type} & \multicolumn{3}{c}{iOS} & \multicolumn{3}{c}{Android} & \multicolumn{19}{c}{Features} & \multicolumn{7}{c}{Personalised services} & \multicolumn{6}{c}{Resources} \\
 & \multirow{2}{*}{1} & \multirow{2}{*}{2} & \multirow{2}{*}{3} & \multirow{2}{*}{4} & \multirow{2}{*}{5} & \multirow{2}{*}{6} & \multirow{2}{*}{7} & \multirow{2}{*}{8} & \multicolumn{12}{c}{Log /Diary} & \multirow{2}{*}{21} & \multirow{2}{*}{22} & \multirow{2}{*}{23} & \multirow{2}{*}{24} & \multirow{2}{*}{25} & \multirow{2}{*}{26} & \multirow{2}{*}{27} & \multirow{2}{*}{28} & \multirow{2}{*}{29} & \multirow{2}{*}{30} & \multirow{2}{*}{31} & \multirow{2}{*}{32} & \multirow{2}{*}{33} & \multirow{2}{*}{34} & \multirow{2}{*}{35} & \multirow{2}{*}{36} & \multirow{2}{*}{37} & \multirow{2}{*}{38} & \multirow{2}{*}{39} & \multirow{2}{*}{40} \\ \cmidrule(lr){10-21}
 &  &  &  &  &  &  &  &  & 9 & 10 & 11 & 12 & 13 & 14 & 15 & 16 & 17 & 18 & 19 & 20 &  &  &  &  &  &  &  &  &  &  &  &  &  &  &  &  &  &  &  &  \\ \midrule
Noom & x & x & 61k & 4.7 & 4 & 311k & 4.2 & 10m & x & x & x &  & x & x & x &  & x &  &  &  & x &  & x & x & x & x & x & x &  &  &  &  &  & Low & x &  &  &  & x &  \\
WeightWatchers & x & x & 226k & 4.8 & 5 & 582k & 4.4 & 10m & x & x & x & x & x & x &  &  & x &  &  &  &  & x & x & x & x &  &  & x & x &  &  &  &  & Medium & x &  &  &  &  &  \\
Lose it! & x & x & 36k & 4.7 & 8 & 104k & 4.2 & 10m & x & x & x & x & x & x & x & x & x & x &  &  &  & x & x &  & x & x & x & x &  &  &  &  &  & Low & x &  &  &  &  &  \\
Slimming world & x & x & 276k & 4.8 & 1 & 17.7k & 4.8 & 1m & x &  &  &  &  &  &  &  & x &  &  &  &  & x & x &  & x & x &  & x &  &  &  &  &  & Low & x &  & x & x &  & x \\
Oviva & x & x & 11k & 4.6 & 4 & 15.6k & 4.5 & 100k & x &  &  &  & x & x &  &  & x &  &  &  &  &  &  & x &  & x &  & x &  & x &  &  &  & Medium & x & x &  &  &  &  \\
MyFitnessPal & x & x & 384k & 4.7 & 19 & 2.72m & 4.5 & 100m & x & x & x & x & x & x & x & x & x &  &  &  & x &  & x & x & x & x &  & x &  &  &  &  &  & Low & x &  & x &  &  &  \\
Yazio & x & x & 7.6k & 4.7 & 19 & 533k & 4.2 & 10m & x & x & x & x & x & x & x & x & x &  &  & x & x & x & x & x &  & x &  & x & x &  & x &  &  & Medium & x &  &  &  &  &  \\
Zero & x & x & 49k & 4.8 & 1 & 60.6k & 4.2 & 1m & x & x &  &  & x &  &  & x & x & x & x &  & x & x &  & x & x &  &  &  &  & x & x & x &  & Medium & x &  &  &  & x &  \\
Healthi & x & x & 831 & 4.7 & 1 & 9.11k & 4.3 & 1m & x &  & x & x &  &  & x &  &  &  &  &  &  &  & x & x & x & x & x & x &  &  &  &  & x & Medium & x &  &  &  &  &  \\
Keto diet app - low carb manager & x & x & 60k & 4.6 & 7 & 8.64k & 4.5 & 500k & x &  & x & x & x & x &  & x & x &  &  &  & x &  & x &  &  & x & x & x &  &  & x &  &  & Medium & x &  &  &  & x &  \\
Carb Manager - Keto Diet Tracker & x & x & 29k & 4.8 & 1 & 144k & 4.7 & 5m & x & x & x & x & x & x & x & x & x &  &  &  &  &  &  &  &  &  &  & x &  &  & x & x &  & Medium & x &  & x & x & x &  \\
Calorie Counter &  & x & 203k & 4.8 & 2 & 43.4k & 4.7 & 1m & x &  & x & x & x & x & x &  & x &  &  & x & x & x & x &  & x & x & x &  &  &  & x &  & x & Medium & x & x & x &  &  & x \\
Monitor your weight & x & x & 20.3k & 4.8 & 23 & 199k & 4.7 & 10m & x &  &  &  &  &  &  &  &  &  &  &  &  &  &  &  &  & x & x &  &  &  & x &  & x & Medium &  &  &  &  &  &  \\
Daily Yoga: Fitness + Meditation &  & x & 148 & 4.4 & 10 & 151k & 4.1 & 10m &  &  &  &  &  &  &  &  & x &  &  &  & x & x &  &  & x &  &  &  & x & x &  &  &  & Medium &  & x & x & x &  &  \\
DietSensor & x & x & 25 & 4.6 & 6 & 347 & 4.5 & 50k & x &  & x & x & x & x & x &  & x &  &  &  &  &  &  & x &  & x &  & x &  &  & x & x & x & High & x &  &  &  &  & x \\
Fastic: Fasting\&Food Tracker & x & x & 62k & 4.8 & 6 & 410k & 4.6 & 10m & x &  & x & x & x & x & x &  &  & x &  &  & x & x & x & x & x & x &  & x & x & x & x &  & x & High & x & x & x &  & x & x \\
Lifesum Food and Calorie tracker & x & x & 27k & 4.6 & 11 & 343k & 4.3 & 10m & x &  & x & x & x & x &  & x & x &  &  &  &  &  & x &  &  & x & x & x &  &  & x & x & x & High & x &  &  &  &  &  \\ \bottomrule
\end{tabular}
\begin{tablenotes}[para,flushleft]
 \scriptsize
1=Web, 2=Mobile, 3=Ratings count, 4=Average rating, 5=Languages, 6=Ratings count, 7=Average rating, 8=Downloads, 9=Weight/BMI, 10=Blood sugar/diabetes, 11=Calories, 12=Macros, 13=Meal, 14=Water in-take, 15=Walking steps, 16=Intermittent fasting, 17=Activity/Exercise, 18=Sleep, 19=Stress, 20=Notes, 21=Reminder or Notification, 22=Achievements or Reward system, 23=Food photo upload or Barcode scanner, 24=Coach or Practitioner, 25=Community Support or Sharing, 26=Synchronise with tracking app, 27=Synchronise with sensing app, 28=Food or recipe, 29=Exercise, 30=Guidance or tips, 31=Statistics (diet or body), 32=Fasting plan, 33=Weigh loss plan or health nutrition target, 34=Recommendation Level, 35=Recipes/Food database/Nutrition, 36=Healthy lifestyle, 37=Activities videos/audio, 38=Podcast, 39=Lessons/Courses, 40=Success story; \\"x" indicate available; “k” = 000, and “m” = 000,000; Low = 1 recommendation; Medium = 2 to 3 recommendations; and High = 4 or more recommendations
\end{tablenotes}
\end{threeparttable}
}
\end{sidewaystable}

\subsection{User needs analysis}
\label{una}
In this study, 207 clients from a UK-based weight management company participated and 121 (20\% Male, 78\% Female and 2\% preferred not to say) completed our survey. 16\% of them completed their weight management programme, while 84\% attended 6-12 weekly sessions. The analysis results of the responses of the participants suggest that they were slightly comfortable with the use of pen and paper (Mean =3.6 (Neutral=3), S.D.=1.18) and smartphone apps (Mean =3.5, S.D.=1.3), and have neutral attitude towards fitness trackers (Mean =3.32, S.D.=1.18), smart sensors (Mean =3.4, S.D.=1.3), and website platform (Mean =3.35, S.D.=1.24) for weight management progress tracking. However, they are uncomfortable (Mean =2.5, S.D.=1.16) with virtual reality headsets for weight management. With regards to resources, participants perceived regular check-ins with programme advisors (Mean =3.59, S.D.=1.65), regular or weekly weight measurement (Mean =3.65, S.D.=1.47), SMART goals (Mean =3.54, S.D.=1.3), and keeping a food diary (Mean =3.38, S.D.=1.43) as helpful resources. However, behaviour change resource (Mean =2.69, S.D.=1.83), fluid intake diary (Mean =2.04, S.D.=1.84), physical activity resources (Mean =2.84, S.D.=1.68), food recipe (Mean =2.66, S.D.=1.84), healthy lifestyle resource (Mean =2.67, S.D.=176), food database (Mean =2.39, S.D.=1.98), and notifications (Mean =2.04, S.D.=1.86) were considered as resources with limited helpfulness. Even worse, weight loss medication (Mean =1.11, S.D.=1.65), thought diary (Mean =1.9, S.D.=1.79), and access to Facebook groups (Mean =1.64, S.D.=1.81) were perceived as the unhelpful resources.

We further applied the t-test to explore whether participants of different sexes, education levels (50\% had higher education), living situations (50\% were living with their spouse or partner), and reasons for joining weight management (73\% of the participants joined a weight management programme for medical reasons, while others for non-medical issues) differed in how comfortable they were using various technology, and how helpful they perceived the weight management resources. The t-test result suggests that compared to males, female participants are significantly more comfortable in using pen and paper (Male: Mean= 3.04, S.D.= 1.2; Female: Mean= 3.73, S.D.= 1.15;\textit{t}(116)=-2.55, \textit{p}<0.05) for weight management progress tracking. Moreover, female perceived significantly more positive in the helpfulness of weight loss medication (Male: Mean=0.5, S.D.= 1.1; Female: Mean= 1.22, S.D.= 1.74;\textit{t}(116)= -2.51, \textit{p}<0.05), and keeping diaries for food (Male: Mean= 2.75, S.D.=1.68; Female: Mean= 3.55, S.D.= 1.34;\textit{t}(116)= -2.18, \textit{p}<0.05), fluid (Male: Mean=1.33, S.D.=1.66; Female: Mean= 2.2, S.D.= 1.88;\textit{t}(116)= -2.23, \textit{p}<0.05) and thought (Male: Mean=1.21, S.D.=1.64; Female: Mean= 2.11, S.D.= 1.82;\textit{t}(116)= -2.34, \textit{p}<0.05). 

Furthermore, participants who joined the weight management programme for medical reasons significantly perceive the helpfulness of keeping a fluid diary as higher than those for non-medical ones (Medical: Mean=2.27, S.D.=1.88; Non-medical: Mean= 1.42, S.D.= 1.6;\textit{t}(119)= 2.47, p<0.05). Finally, participants with higher education significantly felt more comfortable than those with further education in using smartphone apps (Higher education: Mean= 3.67, S.D.= 1.22; Further education: Mean= 3.07, S.D.= 1.4;\textit{t}(99)= 2.2, \textit{p}<0.05), and using pen and paper(Higher education: Mean= 3.8, S.D.= 1.02; Further education: Mean= 3.20, S.D.= 1.35;\textit{t}(99)= 2.47, \textit{p}<0.05), for tracking weight management progress. 

\section{Discussion}
Our review of the commercial DWMIs showed that collected users' data ranged from health and medication conditions, diet, activity, and fitness to individuals' profiles and lifestyles. Based on the nature of collected data, DWMI developers must take appropriate legal and ethical considerations for the collection of users' sensitive information. On the other hand, commercial DWMIs implement essential features, such as logging capabilities, personalised goal settings, synchronisation with external devices and digital community or coaching support etc. These findings align with established literature on weight management \cite{Naabi}, which emphasises the importance of self-monitoring, tailored interventions and social accountability in weight management. The user needs analysis further confirms the significance of some aspects, as participants found features like regular or weekly weight measurement, and SMART goals helpful for weight management. 

Despite the strength of commercial DWMIs, WMI clients perceived some of its established features and services such as food recipes, notification and achievement system, fluid intake diary, and physical activity resources as unhelpful. This indicates a need for more effective design and implementation of these features. Moreover, female participants and those who enrolled in weight management for medical reasons perceived fluid intake and thought diaries as particularly useful, yet a thought diary is rarely implemented into commercial DWMIs. Only 4/17 DWMIs provided notes as a diary feature while none provided mental resources or recommendations. This suggests that commercial DWMIs need to drive the improvement in the use and helpfulness of these features and services among users.

Another area requiring improvement is the personalisation of recommendations. Personalised and adaptive feedback is known to enhance adherence and long-term success in weight management \cite{Ghelani}, however, only four DWMIs offer high-level tailored guidance (Table \ref{table1}, number 34), others provide only a limited number of recommendation types and fail to adapt dynamically to users' progress. Also, findings suggest that while emerging technologies such as artificial intelligence and gamification have the potential to enhance user engagement, they remain underutilised in commercial DMWIs. Although survey participants expressed discomfort with virtual reality headsets, research indicates that AI-driven coaching and game-based incentives can improve motivation and adherence \cite{Spinean}. Future digital weight management solutions could benefit from incorporating these technologies in a user-friendly and accessible manner. Overall, these findings suggest that while commercial DWMIs perform well in areas such as logging or tracking, goal setting and resource provisions, they fall short in supporting behavioural change, adaptive personalisation, and psychological well-being. Future DWMIs should integrate evidence-based behavioural strategies, dynamic goal adjustments, and AI-driven support to better address the complex needs of users.

\section{Conclusion}
This study aimed to identify unique and common features of DWMIs and directions for future design and implementation of digital intervention in weight management by conducting a systematic review of commercial DWMIs and a survey of user needs and their analysis that involved real-life WMI. DWMIs are diverse in terms of functionality, personalisation, and integration with other tracking and sensing applications. Fitness trackers and smartphone-based solutions are the user-preferred digital approaches for tracking weight management progress. DWMIs are most effective when they balance self-monitoring, personalised coaching and behaviour interventions. These factors play key roles in user adoption and effectiveness of DWMI. Understanding these variations helps contextualise their role in digital health and weight management strategies and will also enable users to make informed selection choices.

This study encountered various limitations, such as identifying and including only trending and popular DWMIs via Reddit, reviewing free and not premium versions of DWMIs in only two sessions, manual extraction of information about DWMI services and features, and surveying user needs of only one UK company clients which limit the generalisation of analysis results. In the future, advanced technologies for weight management require further investigation to optimise engagement. Valuable insights into user needs were provided, however, there is a need to explore the longitudinal adherence and real-world efficacy of perceived helpful resources. Also, research into exploring adaptive personalisation for user experience enhancement and long-term adherence should be explored.

%\subsubsection{Acknowledgements} Please place your acknowledgments at
%the end of the paper, preceded by an unnumbered run-in heading (i.e.
%3rd-level heading).

 \bibliographystyle{splncs04}

\end{document}